\documentclass[final,3p,twocolumn,12pt]{elsarticle}

\usepackage{graphicx}

\usepackage{ifpdf}
\ifpdf
\usepackage[dvips]{hyperref}
\else
\usepackage[hypertex]{hyperref}
\fi

\usepackage[T1]{fontenc}

\usepackage{amsmath}
\usepackage{color}

\newcommand{\bra}[1]{{\langle{#1}|}}
\newcommand{\ket}[1]{{|{#1}\rangle}}
\newcommand{\sca}[2]{\langle{#1}|{#2}\rangle}
\newcommand{\ve}[1]{{\overrightarrow{#1}}}
\newcommand{\Tr}{\mathrm{Tr}}
\newcommand{\scal}[2]{\left\langle #1\arrowvert #2\right\rangle}
\newcommand{\vv}[1]{\overrightarrow{#1}}

\newcommand{\ds}{\displaystyle}

\usepackage{amssymb}

\journal{Physica E}

\begin{document}

\begin{frontmatter}



\title{Geometric phase and Pancharatnam phase induced by light wave polarization}


\author[utinam]{J. Lages}
\ead{jose.lages@utinam.cnrs.fr}
\author[femto]{R. Giust}
\author[utinam]{J.-M. Vigoureux}

\address[utinam]{Equipe PhAs, Physique Théorique et Astrophysique, Institut UTINAM, Observatoire des Sciences de l'Univers THETA,
              CNRS \& Université de Franche-Comté, F-25030 Besançon, France}
\address[femto]{Institut FEMTO-ST, UMR CNRS 6174, Université de Franche-Comté,
           16 Route de Gray, 25030 Besançon Cedex, France}

\begin{abstract}
We use the quantum kinematic approach to revisit geometric phases associated with polarizing processes of a monochromatic light wave.
We give the expressions of geometric phases for any, unitary or non-unitary, cyclic or non-cyclic transformations of the light wave state.
Contrarily to the usually considered case of absorbing polarizers, we found that a light wave passing through a polarizer may acquire in general a non zero geometric phase.
This geometric phase exists
despite the fact that initial and final polarization states are in phase according to the Pancharatnam criterion  and can not be measured using interferometric superposition.
Consequently, there is a difference between the Pancharatnam phase and the complete geometric phase acquired by a light wave passing through a polarizer.
We illustrate our work with the particular example of total reflection based polarizers.
\end{abstract}


\end{frontmatter}


\section{Introduction}

The concept of geometric phase naturally arises for polarized light in optics. In 1956, Pancharatnam \cite{pancharatnam56}
studied how the phase of polarized light changes after a cyclic evolution
of its polarization. He found that light wave acquires, in addition to the usual phase associated with the optical path,
a geometric phase depending only on the relative \textit{loci} of the polarization states on the Poincar\'e sphere.
Later, Berry \cite{berry84} developed the concept of geometric phase for dynamical quantum systems with cyclic adiabatic unitary evolutions
and showed its similarity with the Pancharatnam phase  in optics \cite{berry87}. The existence of geometric phase has been also
demonstrated for non unitary and non cyclic evolutions  \cite{aharonov87,samuel88}, and recently for open quantum systems \cite{uhlmann86,uhlmann95,sjoqvist00,viennot11,viennot12}.
Many experiments \cite{bhandari88a,chyba88,bhandari88b,simon88,bhandari93,berry96,hariharan97a,hariharan97b} have provided evidence
for geometric phase in the context of polarized light.
Along a given path, closed or not, on the Poincar\'e sphere, bringing the polarization from a state $\ket 1$ to a state $\ket 2$,
the light wave state acquires a geometric phase which is equal to minus half of the solid angle enclosed by the effectively followed path
and the geodesic connecting states $\ket 1$ and $\ket 2$ \cite{samuel88}. If the path coincides with the geodesic then no geometric phase is gained.
Since any cyclic path on the Poincar\'e sphere is at least a concatenation of two geodesics and is therefore by itself not a geodesic,
a light wave along such a path acquires \textit{de facto} a non zero geometric phase. This property has been widely used in the above cited
experiments where cyclic evolution of the polarization state was usually achieved using retarders (unitary transformations) \cite{bhandari88a,chyba88,bhandari88b,simon88,berry96},
polarizers (non-unitary transformations) \cite{berry96,hariharan97a,hariharan97b},
and both \cite{bhandari88a,bhandari93}.

When a retarder (\textit{e.g.} a wave plate) is used on a light wave, its polarization follows then a piece of
circle on the Poincaré sphere and a geometric phase is consequently acquired (except if the piece of circle is a piece of great circle of length less than $\pi$).
As far as we know, in the literature, the action of a polarizer is considered to not introduce geometric phase 
since it is considered to project the light wave polarization from a state onto another following a geodesic
\cite{bhandari88a,bhandari93,berry96,hariharan97a,hariharan97b}.
This is effectively true for the case of absorbing polarizers. We find that this is no longer true
if one considers total reflection based polarizers since the path followed by the polarization is no more a geodesic but a loxodrome.
More generally, we show that even if a light wave state $\ket1$ is projected onto another state $\ket2\sca{2}{1}$, for example by means of a polarizer, a non zero geometric phase can be acquired by the light wave. At the end of such a transformation the acquired geometric phase is exactly compensated by the acquired dynamic phase in such a way that the total phase has no memory of these two phases.
As a consequence interferometry measurements are not able to capture possible geometric phase acquired during a state projection.

However the Pancharatnam phase \cite{pancharatnam56}, which is a kind of geometric phase, can be measured using interferometry experiments.
As reminded by de Vito and Levrero \cite{devito94}, we can consider that a light wave acquires a Pancharatnam phase
if in the Hilbert space the light wave state is projected successively onto states the polarizations of which marked out a cyclic path on the Poincaré sphere.
The use of successive polarizers ensures that the light wave state is successively projected and then that a Pancharatnam phase is indeed measured \cite{berry96,hariharan97a,hariharan97b}.
We show that the Pancharatnam phase is not the complete geometric phase indeed acquired by a light wave since the Pancharatnam phase does not take into account the geometric phase possibly acquired during the projection processes.

In this paper, we also provide the most general expression of the geometric phase acquired by a light wave experiencing a polarizing transformation. We address particularly the case of the possible geometric phase acquired by a light wave passing through a total reflection based polarizer. 
The paper is organized as follows: 
In Section~\ref{polar}, we define the mathematical formalism describing the light wave state and its polarization. Using the quantum kinematic approach \cite{mukunda93}, in Section~\ref{gauge}, we describe the
geometric phase and the modulus of the degree of coherence as the gauge invariant quantities associated with a non-unitary evolution of the light wave state.
In Section~\ref{device}, we classify any light wave state transformation induced by a polarizing element in terms of $SL(2,\mathbb{C})$ transformations.
In Section~\ref{geom},
first, we revisit the case of unitary transformations corresponding to retarders or to media with optical activity
and the case of non unitary transformation corresponding to absorbing polarizing elements. Then, we show that, in the case of non unitary transformation 
combining both differential attenuation and differential dephasing
a light wave does acquire a geometric phase and we derive its expression.
In Section~\ref{sectph}, we derive for the sake of completeness the total phase and the dynamic phase acquired by a light wave passing through a polarizing device using the model \cite{lages08}. In the frame of this model we express the Pancharatnam criterion \cite{pancharatnam56}.
In Section~\ref{gpp}, we apply the results derived in Section~\ref{geom} for general polarizing elements to the specific case of polarizers. We found that a polarizer in general does induce a non trivial geometric phase and we derive its expression for the case of a total reflection based polarizer.
This geometric phase
exists despite the fact that the initial and final polarization states are \textit{in phase} according to the Pancharatnam criterion and
is a direct reminiscence of the evanescent
component of the
electromagnetic field inside the polarizer.
In the limit where differential absorption is predominant over birefringence (\textit{e.g.} in polaroid films), we retrieve as expected a zero geometric phase. In Section~\ref{sec:difppg}, we discuss the difference between the Pancharatnam phase and the geometric phase acquired by a light wave.



\section{Polarization, space of rays, and Poincar\'e sphere}\label{polar}

A polarized light wave may be described by a vector $\ket \psi$
lying in a two dimensional complex Hilbert space $\cal H$.
Such a vector $\ket \psi$ may be written as
\begin{equation}\label{psi}
\ket\psi =\sqrt I\,e^{i\Phi}\left(\cos\frac \theta 2 \ket 0+e^{i\phi}\sin
\frac \theta 2 \ket 1\right)
\end{equation}
where $I=\scal{\psi}{\psi}\in\mathbb{R}^+$ is the light wave intensity, $\Phi\in[0,2\pi[$ a global phase,
$\phi\in[0,2\pi[$ a relative phase, $\theta\in [0,\pi]$ the polar angle, and
$\left\{\ket0,\ket1\right\}$ an orthonormal basis of $\mathcal{H}$. The vectors $\ket0$ and $\ket1$ represent
{\it e.g.} the normalized state with circular right-handed polarization
and that with circular left-handed polarization respectively.
 
The polarization of the light wave $\ket\psi$ depends only on the ellipticity angle $\chi$ and on the azimuthal angle $\Psi$  \cite{born64}
which are directly related to the polar angle $\theta=\pi/2-2\chi$ and to the relative phase $\phi=2\Psi$.
So,
two light waves $\ket\psi$ and $\ket{\psi'}=a\ket\psi$, where $a$ is a complex factor, share the same polarization. We say
that $\ket{\psi}$ and $\ket{\psi'}$ are equivalent, \textit{i.e.} $\ket{\psi'}\sim\ket{\psi}$, in the sense that it is possible
to convert one of these wave to the other by using a complex scale transformation.
Let us then define the space $\mathcal R$ of unit rays by
$
\mathcal{R}={\cal H}/\sim\,\,=\{\rho=I^{-1}\ket\psi\bra\psi\,\,\,\,|\,\,\ket\psi\in{\cal
H}\}.
$
An element $\rho$ belonging to $\cal R$ may be written as
\begin{equation}\label{proj}
\rho=\frac{1}{2}\left(\sigma^0+\vv
S\cdot\ve \sigma\right)\equiv\rho_{\ve S}
\end{equation}
where $\vv\sigma$ is a three dimensional vector whose components
are the Pauli matrices $\left\{\sigma^i\right\}_{i=1,2,3}$ and where
$\sigma^0$ is the $2\times2$ identity matrix. Any projector $\rho$
is associated with a unique normalized Stokes vector $\ve S=\sin\theta\cos\phi\vv e^1+\sin\theta\sin\phi\ve
e^2+\cos\theta\ve e^3$. The set of the endpoints of all the
normalized Stokes vectors defines the Poincar\'e sphere ${\cal S}^2$. Each $\ve S$ vector is in bijective relation
with a point in the space of rays $\cal R$, \textit{i.e.} with a
projector belonging to $\cal R$. So, the unit Poincar\'e sphere
${\cal S}^2$ is isomorphic to the space of unit
rays ${\cal S}^2\sim {\cal R}$.
The set of vectors
$\{\ket{\psi'}=a\ket\psi, \,a\in\mathbb{C}\}$ corresponds
to a unique projector $\rho_{\ve S}$ (\ref{proj}) and consequently
corresponds to a unique normalized Stokes vector $\ve S$.
A wave $\ket\psi$ as defined in (\ref{psi}) may be
then represented, \textit{modulo a global complex factor}, by a point in the
space of unit rays $\cal R$ or equivalently by a point on the Poincar\'e sphere ${\cal S}^2$.
The
circular right(left)-handed polarization state $\ket0$ ($\ket1$), corresponds to $\theta=0$ ($\theta=\pi$), \textit{i.e.}
to the north (south) pole of the Poincar\'e sphere. Linear polarization states correspond to
vectors of $\mathcal{H}$ with $\theta=\pi/2$, or equivalently correspond to points of the Poincar\'e sphere equator.

\section{Local gauge invariance}
\label{gauge}

Passing through an optical device, the state $\ket\psi$ of a light wave evolves in the Hilbert space $\mathcal{H}$
along a curve
$
\mathcal{C}=\left\{\ket{\psi(s)}\in\mathcal{H}\,\,\arrowvert\,\,s\in\left[s_1,s_2\right]\subset\mathbb{R}\right\}\subset\mathcal{H}.
$
Let us now define another curve
$\mathcal{C}'$ the elements of which are related to the elements of $\mathcal{C}$ by a \textit{local gauge transformation}, $\ket{\psi'(s)}=a(s)\ket{\psi(s)}$.
Here, $a(s)$ is a smooth nonzero complex function of $s\in\left[s_1,s_2\right]$. 
Comparing $\bra{\psi'(s)}\frac{d}{ds}\ket{\psi'(s)}$ with $\bra{\psi(s)}\frac{d}{ds}\ket{\psi(s)}$, it is possible to construct the following complex gauge invariant expression  \cite{mukunda93}
\begin{equation}\label{gi}
\displaystyle\frac{\scal{\psi(s_1)}{\psi(s_2)}}{\scal{\psi(s_1)}{\psi(s_1)}}
\exp\left(-\int_{s_1}^{s_2}ds\frac{\sca{\psi(s)}{\dot\psi(s)}}{\scal{\psi(s)}{\psi(s)}}\right).
\end{equation}
Here the dot denotes the differentiation with respect to the parameter $s$.
Let us define the projection map $\pi:\mathcal{H}\rightarrow\mathcal{R}$ such as, for all $a\in\mathbb{C}$, $\pi(a\ket{\psi})=\pi(\ket{\psi})=\rho\in\mathcal{R}$. Since the curve $\mathcal{C}'$ and $\mathcal{C}$ are related by a gauge transformation,
$\mathcal{C}\sim\mathcal{C}'$, they share the same projected curve image $\mathrm{C}=\pi(\mathcal{C})=\pi(\mathcal{C}')$ in the space of unit rays $\mathcal{R}$.
As expression (\ref{gi}) is gauge invariant, it is a functional of the curve $\mathrm{C}$ and, its modulus $\iota_{\mathrm{g}}[\mathrm{C}]$ and complex argument $\phi_{\mathrm{g}}[\mathrm{C}]$ are also gauge invariant functionals of the curve $\mathrm{C}$. 
The modulus of (\ref{gi}) can be written in the following form
\begin{equation}
\iota_{\mathrm{g}}[\mathrm{C}]=\frac{\left\arrowvert\scal{\psi(s_1)}{\psi(s_2)}\right\arrowvert}{\sqrt{I(s_1)I(s_2)}}=\sqrt{\Tr\left(\rho(s_1)\rho(s_2)\right)}
\end{equation}
which is also the modulus of the complex degree of coherence $\gamma_{12}(0)=\iota_{\mathrm{g}}[\mathrm{C}]e^{i\arg\scal{\psi(s_1)}{\psi(s_2)}}$. Hence
the modulus of the interference term between the two normalized wave states $\left(1/\sqrt{I(s_1)}\right)\ket{\psi(s_1)}$ and $\left(1/\sqrt{I(s_2)}\right)\ket{\psi(s_2)}$ is
a geometric invariant.
The complex argument of (\ref{gi}) is the geometric phase \cite{mukunda93} associated with the curve $\mathrm{C}\subset\mathcal{R}$ 
\begin{equation}\label{pg}
\begin{array}{lll}
\phi_{\mathrm{g}}[\mathrm{C}]&=&\arg{\scal{\psi(s_1)}{\psi(s_2)}}\\
&&-\mathrm{Im}\displaystyle\int_{s_1}^{s_2}ds
\displaystyle\frac{\sca{\psi(s)}{\dot\psi(s)}}{\scal{\psi(s)}{\psi(s)}}\\
&\equiv&\phi_{\,\mathrm{t}}[\mathcal{C}]-\phi_{\mathrm{d}}[\mathcal{C}].
\end{array}
\end{equation}
Here,
\begin{equation}\label{phip}
\phi_{\,\mathrm{t}}[\mathcal{C}]\equiv\arg{\scal{\psi(s_1)}{\psi(s_2)}}
\end{equation}
is the total phase of the curve $\mathcal{C}$, \textit{i.e.} the relative phase of the ending point of $\mathcal{C}$ with respect to its starting point, and 
\begin{equation}\label{phid}
\phi_{\mathrm{d}}[\mathcal{C}]\equiv\mathrm{Im}\displaystyle\int_{s_1}^{s_2}ds
\displaystyle\frac{\sca{\psi(s)}{\dot\psi(s)}}{\scal{\psi(s)}{\psi(s)}}
\end{equation}
is the dynamic phase.
Although $\phi_{\,\mathrm{t}}[\mathcal{C}]$ and $\phi_{\mathrm{d}}[\mathcal{C}]$ are functionals depending on the Hilbert space curve $\mathcal{C}$, their difference $\phi_{\mathrm{g}}[\mathrm{C}]$
is a functional depending only on the corresponding projected curve $\mathrm{C}$ in the unit rays space, and thus is a geometric invariant.
Hence, equivalent trajectories in the Hilbert space, each one related to the other ones by local complex scale transformations, share the same geometric interference term $\iota_{\mathrm{g}}[\mathrm{C}]$ and the same geometric phase $\phi_{\mathrm{g}}[\mathrm{C}]$.


\section{Polarizing devices}
\label{device}

The most general operator affecting the polarization of a given state (\ref{psi}) is proportional
to the $SL(2,\mathbb{C})$ group operator (see \textit{e.g.} \cite{misner73,halpern68})
\begin{equation}\label{sl2C}
\exp\left(\left(\gamma\vv{p_1}-i\delta\vv{p_2}\right)\cdot\vv\sigma/2\right)
\end{equation}
where $\gamma$ and $\delta$ are real and where $\vv{p_1}$ and $\vv{p_2}$ are unit vectors associated with two particular polarization states. In the following, we will consider particular cases of this general operator: (a) the $SL(2,\mathbb C)$ rotation operator $e^{-i\delta/2\vv p\cdot\vv\sigma}$,
(b) the $SL(2,\mathbb C)$ boost operator $e^{\gamma/2\vv p\cdot\vv\sigma}$, and (c) the boost-rotation operator $e^{\left(\gamma-i\delta\right)/2\vv p\cdot\vv\sigma}$.
With the same procedure we use throughout this paper, it is also possible to compute geometric phase for the most general case (\ref{sl2C}) where vectors $\vv{p_1}$ and $\vv{p_2}$ are non collinear, but the result is hardly understandable in simple geometrical terms such as area decomposition on the Poincaré sphere. Moreover, as far as we know, this case does not correspond to an elementary light wave transformation, contrary to the particular cases (a), (b) and (c) \cite{jones42,pancharatnam55,liu96}.

Let us consider a light wave initially in the state $\ket{\psi(0)}$ passing through an optical device and transformed into
\begin{equation}\label{tf}
\ket{\psi(z)}=a(z)\,e^{\frac{A(z)}{2}\vv p\cdot\vv\sigma}\ket{\psi(0)}.
\end{equation}
Here, the evolution parameter is $z$ which can be identified to the penetration length of the light into the optical device.
The function $a(z)=|a(z)|e^{i\alpha(z)}$ put together global light wave attenuation $|a(z)|$ and global propagation phase $\alpha(z)$. By definition (\ref{tf}), $|a(0)|=1$
and $\alpha(0)=0$.
Independently of its exact form, the function $a(z)$ constitutes a local gauge degree of freedom since it does not alter the light wave polarization. 
In the case of unitary transformation 
$|a(z)|=1$ for all $z$.
The function $A(z)$, such as $A(0)=0$, determines the type of transformation:
(a) In the case of non absorbing birefringent devices or of media with optical activity, the function $A(z)$ is
imaginary, \textit{i.e} $A(z)=-i\delta(z)$. For birefringent material, $\delta(z)$ is the difference between the ordinary and the extraordinary phases associated with the propagation of the optical field along each optical axis.
The unitary transformation (\ref{tf}) is then
equivalent in the Poincar\'e space
to the rotation
of the light wave polarization $\vv S$ around the axis $\vv p$ by
an angle $\delta(z)$.
(b) In the case of absorbing polarizers, such as polaroid films,
the function $A(z)$ is real, \textit{i.e} $A(z)=\gamma(z)$. The function $\gamma(z)$ is then the difference between the light wave attenuations in the polarizer axis direction $\vv p$ and in its orthogonal direction $-\vv p$.
During the non unitary transformation (\ref{tf}),
the polarization $\vv S$ is progressively brought toward the polarizer axis $\vv p$.
Modulo a global attenuation term $|a(z)|$, the transformation 
(\ref{tf}) can be seen as a Lorentz boost transformation written in the $SL(2,\mathbb{C})$ group representation.
(c) Finally, in the case where the function $A(z)=\gamma(z)-i\delta(z)$ is complex,
the two effects (a) and (b) described above act together on the light wave state.
This is for example the case of total reflection based polarizers \cite{lages08}, \textit{i.e.} polarizers using the fact
that at the interface between a medium with index $n$ and a birefringent medium with indexes $n_o$ and $n_e$, satisfying
$n_o>n>n_e$, only the light wave component associated with $n_o$ is transmitted for suitable incident angles.
The case (c) corresponds also to the case of optical devices which are both absorbing and birefringent. For example, realistic absorbing polarizers present inherently a small amount of birefringence.

\section{Geometric phase induced by polarizing transformations}\label{geom}

Let us now compute the geometric phase $\phi_{\mathrm{g}}[\mathrm{C}]$ for the different types of transformation
(a), (b) and (c) defined in the paragraph above. For that purpose we use the horizontal lift $\mathcal{C}_\mathrm{h}\subset\mathcal{H}$ of the unit
ray space curve $\mathrm{C}\subset\mathcal{R}$. For such a curve $\mathcal{C}_\mathrm{h}$, the dynamic phase vanishes, $\phi_{\mathrm{d}}[\mathcal{C}_\mathrm{h}]=0$, since for any $\ket{\psi(s)}\in\mathcal{C}_{\mathrm{h}}$, with $s\in[0,z]$, we have $\mathrm{Im}\sca{\psi(s)}{\dot\psi(s)}=0$. The geometric phase
(\ref{pg}) reads then 
\begin{equation}\label{pgh}
\phi_{\mathrm{g}}[\mathrm{C}]=\,\arg\sca{\psi(0)}{\psi(z)}.
\end{equation}

\subsection{Unitary case ($\delta\neq0,\gamma=0$)}
\begin{figure}[t]
\begin{center}
\includegraphics[width=0.8\columnwidth]{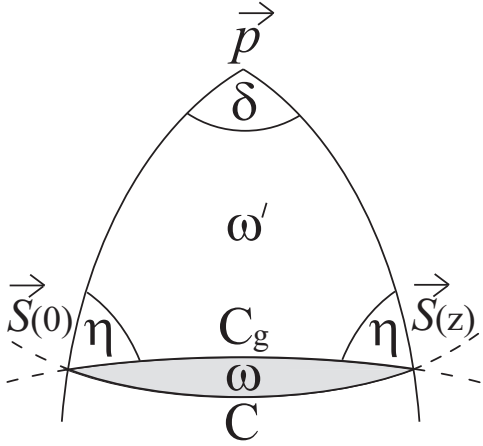}
\caption{\label{fig1}
Polarization trajectory on the Poincar\'e sphere corresponding to an unitary transformation of the light wave state.
The ray space trajectory is $\mathrm{C}$. The curve $\mathrm{C_g}$ is the geodesic connecting the endpoints of $\mathrm{C}$, \textit{i.e.} $\vec S(0)$ and $\vec S(z)$.
During the unitary evolution of the Stokes vector, the accumulated solid angle is $\Omega(z)=\omega'(z)+\omega(z)$.
The geometric phase accumulated along the trajectory $\mathrm{C}$ is minus half of the shaded area,
\textit{i.e.} $\phi_{\mathrm{g}}[\mathrm{C}]=-\displaystyle\frac{\omega(z)}{2}$.
}
\end{center}
\end{figure}

Using (\ref{tf}) for unitary transformations (a), the condition $\ket{\psi(s)}$ belongs to $\mathcal{C}_{\mathrm{h}}$ implies for the global phase $\alpha(z)$ the following relation 
\begin{equation}\label{eq1}
\alpha(z)=-\displaystyle\frac{\Omega(z)}{2}+\frac{\delta(z)}{2}
\end{equation}
where
\begin{equation}\label{solid}
\Omega(z)=\delta(z)(1-\cos\beta)
\end{equation}
is the accumulated solid angle subtended by the
unit Poincar\'e sphere surface swept during the transformation by the arc length joining the endpoint of the normalized Stokes vector $\vv S(z)$ 
and the endpoint of the rotation vector $\vv p$ (see Figure \ref{fig1}).
In (\ref{solid}), $\beta$ is the angle between the Stokes vector $\vv{S}$ and the polarizing device characteristic vector $\vv p$.
Using (\ref{tf}), (\ref{pgh}), and (\ref{eq1}), the calculation of the geometric phase gives
\begin{equation}\label{phigg0}
\begin{array}{lll}
\phi_{\mathrm{g}}[\mathrm{C}]&=&-\frac{\Omega(z)}{2}+\frac{\delta(z)}{2}\\
&&-\tan^{-1}\left(\tan\frac{\delta(z)}{2}\cos\beta
\right).
\end{array}
\end{equation}

For the particular case of a closed loop, \textit{i.e.} a rotation of $\delta(z)=2n\pi$ ($n\in\mathbb{N^*}$), we retrieve the expected result for the geometric phase,
$\phi_{\mathrm{g}}=-n\frac{\Omega_0}2=-n\pi(1-\cos\beta)$, which is minus $n$ times half of the solid angle $\Omega_0$ enclosed by a single loop.

For unclosed loops, the sum of the second and the third terms in (\ref{phigg0}) does not vanish and is equal to
half
of the area
$\omega'(z)$ of the spherical triangle connecting the points $\ve p$, $\ve S(0)$ and $\ve S(z)$ (see Figure \ref{fig1}).
Since, $\Omega(z)=\omega'(z)+\omega(z)$ (see Figure \ref{fig1}), the geometric phase (\ref{phigg0}), $\phi_{\mathrm{g}}[\mathrm{C}]=-\frac{\Omega(z)}{2}+\frac{\omega'(z)}{2}$, is then minus half of the shaded
area $\omega(z)$ enclosed by the ray space trajectory $\mathrm{C}$ and the geodesic $\mathrm{C}_\mathrm{g}$ connecting the endpoints of $\mathrm{C}$,
\begin{equation}
\phi_{\mathrm{g}}[\mathrm{C}]=-\frac{\omega(z)}{2}.
\end{equation}
This result can be easily checked using elementary spherical geometry, $\omega'(z)=\delta(z)+2\eta(z)-\pi$.
The angle $\eta(z)$ in Figure \ref{fig1} can be easily calculated as the angle at the vertex $\vv S(0)$ between the tangent vector associated to the geodesic
$\vv S(0)\rightarrow\vv p$ and the tangent vector associated to the geodesic $\vv S(0)\rightarrow\vv S(z)$. The calculus gives
$\eta(z)=\pi/2-\tan^{-1}\left(\tan\left(\delta(z)/2\right)\cos\beta\right)$.

\subsection{Pure absorption case ($\delta=0,\gamma\neq0$)}
Following the same procedure as above, \textit{i.e.}
we calculate the geometric phase along the horizontal lift $\mathcal{C}_\mathrm{h}$ of $\mathrm{C}$, we obtain trivially for the pure absorbing transformations (b)
\begin{equation}\label{eq3}
\alpha(z)=\alpha(0)=0.
\end{equation}
The geometric phase calculated with (\ref{tf}) and (\ref{eq3}) is then zero,
\begin{equation}\label{phigd0}
\phi_{\mathrm{g}}[\mathrm{C}]=\,0.
\end{equation}
This result is explained by the fact that, for pure absorption, the normalized Stokes vector
describes a geodesic connecting the initial and final polarization states on the unit Poincar\'e sphere \cite{lages08},
and thus the area $\omega(z)$ in Figure \ref{fig1} is zero (the curves $\mathrm{C}$ and $\mathrm{C_g}$ are the same, $\mathrm{C}=\mathrm{C_g}$).

\begin{figure}
\begin{center}
\includegraphics[width=0.9\columnwidth]{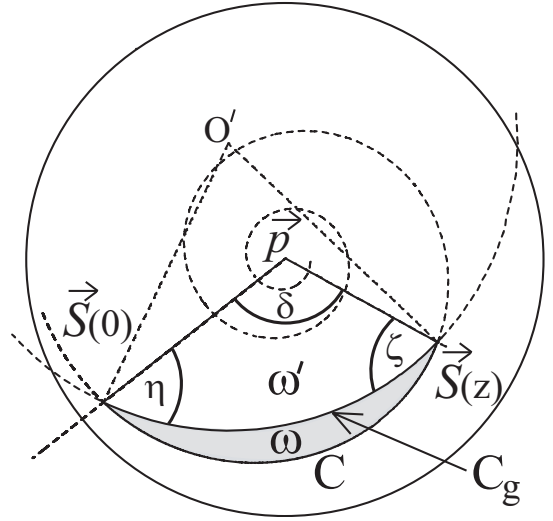}
\caption{\label{fig2}
Polarization trajectory on the Poincar\'e sphere surface corresponding to a non unitary transformation of the light wave state.
We use a stereographic projection of the ray space trajectory viewed from the pole $-\vec p$.
The ray space trajectory $\mathrm{C}$ is a piece of loxodrome \cite{lages08} on the Poincar\'e sphere and its stereographic projection is a piece of logarithmic spiral. The curve $\mathrm{C_g}$ is the geodesic connecting the endpoints, $\vec S(0)$ and $\vec S(z)$, of $\mathrm{C}$, its stereographic projection is a circular arc with center $\mathrm{O}'$.
As in Figure \ref{fig1}, the accumulated solid angle is $\Omega(z)=\omega'(z)+\omega(z)$ and the geometric phase accumulated along the trajectory $\mathrm{C}$ is minus half of the shaded area, \textit{i.e.} $\phi_{\mathrm{g}}[\mathrm{C}]=-\displaystyle\frac{\omega(z)}{2}$.
}
\end{center}
\end{figure}

\subsection{General case ($\delta\neq0,\gamma\neq0$)}
\label{general}

In the general case (c), the calculus of the geometric phase using the horizontal lift $\mathcal{C}_h$ of $\mathrm{C}$
forces $\alpha(z)$, 
to obey the following expression
\begin{equation}\label{aberr}
2\frac{d\alpha}{d\delta}=\frac{\tanh\gamma(s)+\cos\beta(0)}{1+\tanh\gamma(s)\cos\beta(0)}
\end{equation}
where $\beta(0)$ is the angle between the initial light wave polarization $\vv S(0)$ and the vector $\vv p$ associated with the polarizing transformation.
The right-hand side of this expression is analogous to the relativistic formula for aberration of light. Indeed, considering two reference frames $\mathcal{K}$ and $\mathcal{K}'$, the latter moving from the first at the velocity $v$, replacing $\gamma$ by the rapidity $\Phi=\arg\tanh\displaystyle\frac vc$, and finally considering $\beta(0)$ as the angle through which an observer at rest in $\mathcal{K}$ observes a given star, the right hand side of expression (\ref{aberr}) is equal to the cosine of the apparent angle $\beta(s)$ through which an observer at rest in $\mathcal{K}'$ observes the same given star. This analogy allows us to define
\begin{equation}\label{cos}
\begin{array}{ccc}
\beta(s)&=&\beta\circ\gamma(s)\\
&=&\arccos\left(
\ds\frac{\tanh\gamma(s)+\cos\beta(0)}{1+\tanh\gamma(s)\cos\beta(0)}
\right)
\end{array}
\end{equation}
which is \cite{lages08,giust09} the angle between the Stokes vector $\vv S(s)$, representing the light wave polarization at $s\in[0,z]$, and the vector $\vv p$. This angle $\beta(s)$ increases from $\beta(0)$
to $0$ as $\gamma$ increases from $0$ to $+\infty$.

Integrating (\ref{aberr}) with the help of (\ref{cos}), the global phase $\alpha(z)$ can be written as
\begin{equation}\label{eq2}
\alpha(z)=-\displaystyle\frac{\Omega(z)}{2}+\frac{\delta(z)}{2}
\end{equation}
where 
\begin{equation}\label{eq22}
\Omega(z)=
\int_{\Omega(z)}d\Omega
\end{equation}
is the accumulated solid angle $\Omega(z)$.
Here $d\Omega=\sin\beta d\beta d\delta$ is the surface element of the unit Poincar\'e sphere.
Equations (\ref{eq1}-\ref{solid}) and (\ref{eq3}) are particular cases of the more general equation (\ref{eq2}-\ref{eq22}).

Now, using (\ref{tf}), (\ref{pgh}) and (\ref{eq2}), the geometric phase is
\begin{equation}\label{phigg}
\begin{array}{lll}
\phi_{\mathrm{g}}[\mathrm{C}]&=&-\frac{\Omega(z)}{2}+\frac{\delta(z)}{2}\\
&&
-\tan^{-1}\left(\tan\frac{\delta(z)}{2}\cos\beta\circ\displaystyle\frac{\gamma(z)}{2}
\right).
\end{array}
\end{equation}
This expression of the geometric phase is the generalization of the unitary case expression (\ref{phigg0}) ($\gamma=0$) and also trivially of the pure absorption case expression (\ref{phigd0}) ($\delta=0$). As in the unitary case,
the geometric phase (\ref{phigg}) is equal to minus half of the area $\omega(z)$ delimited by the effectively followed unit ray space trajectory $\mathrm{C}$ 
and the unit ray space geodesic $\mathrm{C_g}$ connecting the endpoints of $\mathrm{C}$. In Figure \ref{fig2}, we represent the stereographic projection of $\mathrm{C}$, $\mathrm{C_g}$, and $\omega(z)$ (shadded area).


\section{Total phase and dynamic phase induced by elementary polarizing devices}
\label{sectph}

For the sake of completeness we also discuss in this paragraph the total phase $\phi_{\,\mathrm{t}}[\mathcal{C}]$ and the dynamic phase $\phi_{\mathrm{d}}[\mathcal{C}]$ induced by elementary polarizing devices. Contrary to the geometric phase $\phi_{\mathrm{g}}[\mathrm{C}]$ which depends only on the effectively followed path $\mathrm{C}$ on the Poincaré sphere, the total and the dynamic phases 
depend on the followed path $\mathcal{C}$ in the Hilbert space $\mathcal{H}$. Consequently, we have to consider a given model describing the Hilbert space evolution of the light wave state. In the following, we use the model presented in Ref.~\cite{lages08} to depict a light wave passing through uniaxial elementary optical devices such as wave plates, media with optical activity, absorbing polarizers, and total reflection based polarizers. Within this model \cite{lages08}, the state $\ket{\psi(0)}$ of a light wave passing through such optical devices is transformed according to
\begin{equation}\label{model}
\ket{\psi(z)}=e^{-ik_oz}e^{-\mu_oz}e^{-\frac{\gamma-i\delta}2}e^{\frac{\gamma-i\delta}{2}\vv p\cdot\vv\sigma}\ket{\psi(0)}
\end{equation}
where $\delta(z)=\left(k_o-k_e\right)z$ is the phase difference induced by the optical device between the ordinary and the extraordinary light wave components and $\gamma(z)=\left(\mu_e-\mu_o\right)z$ is the difference between light wave attenuation rates along the two axes.
In the case of wave plates or elements with optical activity, the vector $\vv p$ is the vector about which the polarization $\vv S(0)$ is rotated by an angle $\delta(z)$. In the case of polarizers, the vector $\vv S(0)$ is progressively brought along the polarization states vector $\vv p$.
Using the definitions provided by (\ref{phip}) and (\ref{phid}), the transformation (\ref{model}) gives the following expressions for the total phase
\begin{equation}\label{tph}
\begin{array}{lll}
\phi_{\,\mathrm{t}}[\mathcal{C}]&=&-k_oz+\displaystyle\frac{\delta(z)}{2}\\
&&-\tan^{-1}\left(\tan\displaystyle\frac{\delta(z)}{2}\cos\beta\circ\displaystyle\frac{\gamma(z)}{2}\right)\\
&=&-k_oz+\displaystyle\frac{\omega'(z)}{2}
\end{array}
\end{equation}
and for the dynamic phase
\begin{equation}\label{dph}
\phi_{\mathrm{d}}[\mathcal{C}]=-k_oz+\displaystyle\frac{\Omega(z)}{2}
\end{equation}
where $\omega'(z)$ is the area of the geodesic triangle $(\vv p,\vv S(0),\vv S(z))$ on the Poincaré sphere (see Figures \ref{fig1} and \ref{fig2}) and 
where $\Omega(z)$ is again the accumulated solid angle (\ref{eq22}), \textit{i.e.} the area $\omega+\omega'$ swept by the arc length joining the polarizing vector $\vv p$ and the Stokes vector $\vv S$ during the evolution of the polarization state. We immediately retrieve the expression of the geometric phase $\phi_{\mathrm{g}}[\mathrm{C}]$ (\ref{phigg}) in the general case ($\gamma\neq0,\delta\neq0$) if we consider the difference between (\ref{tph}) and (\ref{dph}).

In equations (\ref{tph}) and (\ref{dph}), the velocity of the ordinary light wave component seems to play a particular role. This is due to the fact that in the model (\ref{model}) the phase $e^{-ik_oz}$ is factorized so the other phase terms depend only on $\delta(z)$ which involves the relative velocity between the two light wave components. In fact none of the two light wave components (ordinary or extraordinary) plays a particular role as equations (\ref{tph}) and (\ref{dph})
can be rewritten in a more symmetric way respectively as
\begin{equation}
\begin{array}{lll}
&&\phi_{\,\mathrm{t}}[\mathcal{C}]=-\displaystyle\frac{\left(k_o+k_e \right)z}{2}\\
&&-\tan^{-1}\left(\tan\displaystyle\frac{\left(k_o-k_e\right)z}{2}\cos\beta\circ\displaystyle\frac{\gamma(z)}{2}\right)
\end{array}
\end{equation}
and as
\begin{equation}
\begin{array}{lll}
\phi_{\mathrm{d}}[\mathcal{C}]&=&-\displaystyle\frac{\left(k_o+k_e \right)z}{2}\\
&&-\ds\frac{\left(k_o-k_e\right)}2\ds\int_0^zds\,\cos\beta\circ\gamma(s).
\end{array}
\end{equation}
In equations (\ref{tph}) and (\ref{dph}), it would have been possible to single out the $-k_ez$ term instead of the $-k_oz$ term, the definition of $\delta$ would have been changed by a minus sign. 

\subsection{Pancharatnam \textit{in phase} criterion}
\label{sec:pancharatnam}

The Pancharatnam criterion \cite{pancharatnam56} states that two light wave states $\ket{\psi_1}$ and $\ket{\psi_2}$ are \textit{in phase} if their interference gives the maximum intensity, \textit{e.g.} if $\arg\scal{\psi_1}{\psi_2}=0$. Although the two light waves experience different polarization processes, the two light wave states are said \textit{in phase} if the polarization processes does not introduced extra phases.
It is important to note that in the original paper \cite{pancharatnam56} only pure polarization processes
are considered, and propagation phases are not considered.

In the model (\ref{model}), once the pure propagation phase term $e^{-ik_oz}$ is factorized, the polarization process is completely determined by $\delta(z)$, $\gamma(z)$, $\vv p$, and $\vv S(0)$. The total phase will be
\begin{equation}
\phi_{\,\mathrm{t}}[\mathcal{C}]=-k_oz+
\begin{array}{c}
\mbox{extra phase}\\
\mbox{due to the}\\
\mbox{polarization process.}
\end{array}
\end{equation}
For a given polarizing device the free evolution term $-k_oz$ is constant since it corresponds to the acquired phase during the free evolution of the light wave through a medium of optical index $n_o$ and fixed depth $z$.
Hence, within the particular model (\ref{model}),
no extra phase coming from the polarization process is added between the initial state $\ket{\psi(0)}$ and the final state $\ket{\psi(z)}$ if, for any orientation of the polarization device,  $\phi_{\,\mathrm{t}}[\mathcal{C}]=\arg\scal{\psi(0)}{\psi(z)}=-k_oz$.
If we want to bring the initial state $\ket{\psi(0)}$ and the final state $\ket{\psi(z)}$ to interfere with each other, we will observe the interference of the light wave states $\ket{\psi_1}=e^{i\delta_1}\ket{\psi(0)}$ and $\ket{\psi_2}=e^{i\delta_2}\ket{\psi(z)}$ where $\delta_1$ and $\delta_2$ are directly related to optical paths. If no extra phase coming from the polarization process is induced by the polarizing device, the intensity of the sum of these two states will be modulated by the cosine of the angle $\arg\scal{\psi_1}{\psi_2}=-k_oz+\delta_2-\delta_1$.
This constant angle can be reduced to zero if optical paths are chosen conveniently, and then the Pancharatnam criterion is fulfilled. We recall again that in the Pancharatnam original paper \cite{pancharatnam56}, only phase changes due to polarizing processes were considered. We will therefore consider, in the following, that the Pancharatnam  \textit{in phase} criterion can be fulfilled once $\phi_{\,\mathrm{t}}[\mathcal{C}]=\arg\scal{\psi(0)}{\psi(z)}=-k_oz$.

\subsection{Wave plates or media with optical activity}

For wave plates or media with optical activity ($\gamma=0$, $\mu_o=0$), the transformation (\ref{model}) is unitary and introduces a non trivial total phase
\begin{equation}\label{tphu}
\begin{array}{lll}
\phi_{\,\mathrm{t}}[\mathcal{C}]&=&-k_oz+\displaystyle\frac{\delta(z)}{2}\\
&&-\tan^{-1}\left(\tan\frac{\delta(z)}{2}\cos\beta\left(0\right)\right)
\end{array}
\end{equation}
and the following dynamic phase
\begin{equation}\label{dphu}
\phi_{\mathrm{d}}[\mathcal{C}]=-k_oz+\frac{\delta(z)}{2}\left(1-\cos\beta(0)\right).
\end{equation}
The difference between the total phase $\phi_{\,\mathrm{t}}[\mathcal{C}]$ and the dynamic phase $\phi_{\mathrm{d}}[\mathcal{C}]$ acquired by a light wave passing through a retarder or a media with optical activity gives back as expected the geometric phase $\phi_{\mathrm{g}}[\mathrm{C}]$ for unitary transformations (\ref{phigg0}).

Experiments \cite{bhandari88a,chyba88,bhandari88b,simon88} using only retarders to cycle the polarization state in order to measure the geo\-me\-tric Pancharatnam phase have been criticized \cite{devito94}
on the ground that two successive light wave states of the cycle are not \textit{in phase} since an extra phase due to the polarization process is introduced between them. This is clearly seen in (\ref{tphu}) where $\phi_{\,\mathrm{t}}[\mathcal{C}]\neq-k_oz$.

\subsection{Polarizers}
\label{sec:pol}

A polarizer is used to convert any light wave polarization into a specific one.
In other words, a polarizer brings continuously any light wave Stokes vector $\vv S$ along a given direction $\vv p$ related with the polarizer axis. In experimental setups, efficient polarizers are required which means that $\gamma(z)\gg 1$. In this limit, Eq.~(\ref{cos}) gives $\cos\beta\circ\frac{\gamma(z)}{2}=1$, so 
the total phase (\ref{tph}) and the dynamic phase (\ref{dph}) accumulated by a light wave through any polarizer are respectively
\begin{equation}\label{eq:phip}
\phi_{\,\mathrm{t}}[\mathcal{C}]=-k_oz
\end{equation}
and
\begin{equation}\label{eq:phid}
\phi_{\mathrm{d}}[\mathcal{C}]=-k_oz+\lim_{\gamma(z)\rightarrow\infty}\frac{\Omega(z)}{2}.
\end{equation}

The result (\ref{eq:phip}) corroborates the fact that a polarizer does not induce any extra phase coming from the polarization process. The Pancharatnam \textit{in phase} criterion can therefore be fulfilled using a light wave going through a polarizer and the same light wave evolving freely. It is important to note that this result holds for any type of polarizer, whether it is mainly absorbing such as polaroid films, or not as for example in the case of total reflexion based polarizers.

The result (\ref{eq:phip}) can also be rapidly deduced using geometric properties. Indeed, $\omega'(z)$ in (\ref{tph}) is the area of the $(\vv p,\vv S(0),\vv S(z))$-geodesic triangle (see Figure \ref{fig2}). For a perfect polarizer it is required that $\vv S(z)=\vv p$, so the area of the geodesic triangle is zero.


\section{Geometric phase induced by polarizers}
\label{gpp}

Using (\ref{eq:phip}) and (\ref{eq:phid}),
the geometric phase acquired by a light wave going through any type of polarizer ($\gamma(z)\gg 1$) is \textit{a priori} non zero and reads
\begin{equation}\label{phigf}
\phi_{\mathrm{g}}[\mathrm{C}]\equiv\phi_{\,\mathrm{t}}[\mathcal{C}]-\phi_{\mathrm{d}}[\mathcal{C}]=-\lim_{\gamma(z)\rightarrow\infty}\frac{\Omega(z)}{2}.
\end{equation}
Expression (\ref{phigf}) is the most general expression for the geometric phase induced by a polarizer.
It is worth to note in the case of polarizers, as the total phase is constant (\ref{eq:phip}), variations in the geometric phase (\ref{phigf}) and variations in the dynamic phase (\ref{eq:phid}) are compensating each other.

Before considering the geometric phases induced by absorption based polarizers (Section~\ref{sec:ap}) and total reflection based polarizers (Section~\ref{trbp}), let us calculate for the general case transformations ($\delta\neq0,\gamma\neq0$)
(see Section~\ref{general}) the expression of the accumulated solid angle $\Omega(z)$ entering the expression of the geometric phase (\ref{phigg}) and consequently (\ref{phigf}). This expression will be useful in the following.
Using (\ref{cos}) and the light wave intensity expression,
\begin{equation}
\ds\frac{I(s)}{I(0)}=\left|a(s)\right|^2\left(\cosh\gamma(s)+\sinh\gamma(s)\cos\beta(0)\right),
\end{equation}
computed from (\ref{tf}),
it is possible to rewrite the accumulated solid angle $\Omega(z)$ defined in (\ref{eq22}) as
\begin{equation}\label{eq4}
\Omega(z)=\delta(z)-\int_0^zds\,\frac{\dot{\delta}(s)}{\dot{\gamma}(s)}\frac{d}{ds}\ln\displaystyle\frac{I(s)}{I(0)\left|a(s)\right|^2}.
\end{equation}
A convenient property of the geometric invariants $\iota_{\mathrm{g}}[\mathrm{C}]$ and $\phi_{\mathrm{g}}[\mathrm{C}]$ is that these functionals are parametrization invariant \cite{mukunda93}.
Hence, choosing freely the parametrization of $\delta(s)$ and $\gamma(s)$ will not affect the geometric phase associated with $\mathrm{C}$. 
We choose the natural and convenient parametrization already used in Section~\ref{sectph} which is, $\gamma(s)=\Gamma s$ and  $\delta(s)=\Delta s$, where $\Gamma=\mu_e-\mu_o$ is the differential attenuation rate of the optical device and $\Delta=k_o-k_e$
the differential propagation rate. The parameter $s$ is then the penetration length inside the polarizing device. Using now this particular parametrization we can easily integrate (\ref{eq4}) as
\begin{equation}\label{eq:Omega}
\begin{array}{l}
\Omega(z)=\Delta z\\
-\ds\frac{\Delta}{\Gamma}\ln
\left(
e^{\Gamma z}\cos^2\ds\frac{\beta(0)}{2}+
e^{-\Gamma z}\sin^2\ds\frac{\beta(0)}{2}
\right).
\end{array}
\end{equation}

\subsection{Absorption based polarizers}
\label{sec:ap}

Absorption based polarizers (dichroic polarizers) absorb a light wave polarization state component more efficiently than its orthogonal light wave polarization state. Although dichroic polarizers present inherently some birefringence due to their anisotropy, their principal characteristic is the dichroism. For such polarizers it is reasonable to assume that the absorption rate $\Gamma$ is much greater than the dephasing rate $\Delta$. Hence assuming $\ds\frac\Delta\Gamma\rightarrow 0$ and $\Gamma\gg1/z$ in (\ref{eq:Omega}) and (\ref{phigf}), we retrieve the fact that in a very good approximation a dichroic polarizer, such as a polaroid film, does not induce any geometric phase since
\begin{equation}\label{eq:phiap}
\phi_{\mathrm{g}}[\mathrm{C}]=0.
\end{equation}
The fact that for such polarizers
\begin{equation}
\lim_{\Delta/\Gamma\rightarrow0}\Omega(z)=0
\end{equation}
clearly illustrates that
the path followed by the light wave state on the Poincaré sphere is the geodesic connecting the initial to the final polarization state.

It is worth to note that no contribution from the polarization process enters the expression of the dynamic phase acquired by a light wave through an absorbing polarizer since
\begin{equation}
\phi_{\mathrm{d}}[\mathcal{C}]=-k_oz.
\end{equation}

\subsection{Total reflexion based polarizers}
\label{trbp}

\begin{figure}[t]
\begin{center}
\includegraphics[width=0.9\columnwidth]{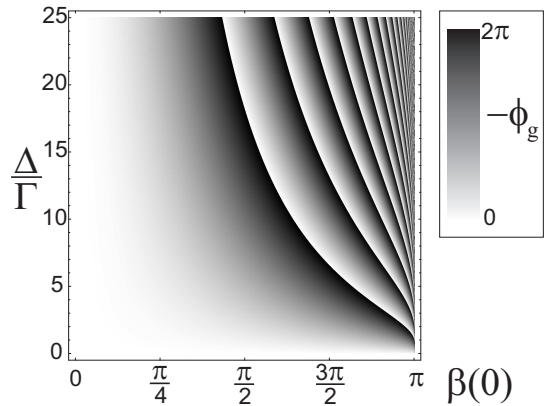}
\caption{\label{fig3}
Geometric phase $\phi_{\mathrm{g}}$ for the case of a total reflection based polarizer.
Density plot of $-\phi_{\mathrm{g}}\mod2\pi$ as a function of $\Delta/\Gamma$ and $\beta(0)$.
}
\end{center}
\end{figure}

In the case of total reflection based polarizers, we have to take into account the fact that $\Delta$ and $\Gamma$ are
characteristics of the same order of magnitude.
Indeed, considering a light wave coming from a medium with an index $n$ and entering at $z=0$ a birefringent medium with indexes $n_o$ and $n_e$, satisfying $n_0>n>n_e$, the continuity of the Maxwell equations at $z=0$ gives $\Delta/\Gamma=\sqrt{\left(n_0^2-n^2\sin^2i\right)/\left(n^2\sin^2i-n_e^2\right)}$ where $i$ is a convenient incident angle ensuring the total reflection of the extraordinary light wave component.
The ratio $\Delta/\Gamma$ is then \textit{a priori} finite.
In those conditions, \textit{i.e.} $\Gamma\gg 1/z$ and $\Delta/\Gamma=cte$,
the accumulated geometric phase (\ref{phigf}) is not trivially zero as in the case of absorption based polarizer (\ref{eq:phiap}), but instead finite
\begin{equation}\label{eq:phigf}
\phi_{\mathrm{g}}[\mathrm{C}]=\frac{\Delta}{\Gamma}\ln\cos\frac{\beta(0)}{2}.
\end{equation}
This geometric phase 
depends only on the distance $\beta(0)$ between 
the initial
and final light wave polarization states, and, on the characteristic ratio $\Delta/\Gamma$ of the polarizer.
Even though the evanescent field component of the light wave dies within few wavelengths just after the interface ($z=0$) and obviously does not
contribute to the final light wave polarization,
information on its attenuation rate $\Gamma$ and its dephasing rate $\Delta$ is nevertheless encoded in the geometric phase (\ref{eq:phigf}). Hence, this geometric
phase is a direct reminiscence of the evanescent field component existing inside a total reflection based polarizer.
Figure \ref{fig3} shows the density plot of the geometric phase $\phi_{\mathrm{g}}[\mathrm{C}]$
as $\beta(0)$ varies from $0$ to $\pi$ and as the ratio $\Delta/\Gamma$ varies from $0$ to $25$.
We note that as the initial distance $\beta(0)$ approaches $\pi$ the geometric phase $\phi_{\mathrm{g}}[\mathrm{C}]$ is rapidly varying.

As even absorption based polarizers, such as polaroid films, are inherently birefringent, the zero induced geometric phase in the case of ideal absorption based polarizer (\ref{eq:phiap})
can be seen as the limit of the geometric phase (\ref{eq:phigf}) where the rate of attenuation/absorption ($\Gamma$) is predominant over the rate of dephasing ($\Delta$) due to birefringence.

\section{Difference between Pancharatnam phase and geometric phase}
\label{sec:difppg}

In this section, we illustrate the difference between the Pancharatnam phase and the geometric phase acquired by a light wave.

\begin{figure}[t]
\begin{center}
\includegraphics[width=0.9\columnwidth]{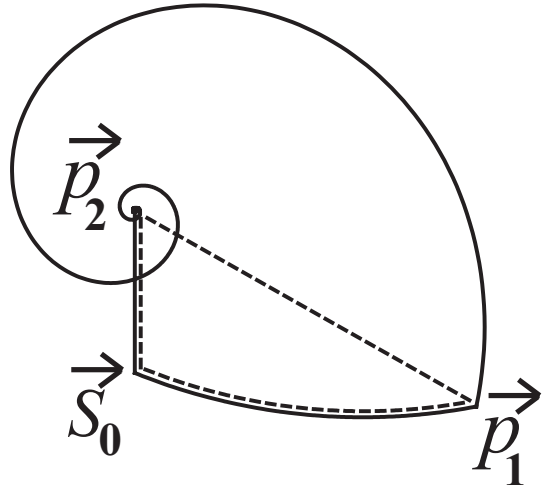}
\caption{\label{fig4}
Polarization trajectory on the Poincaré sphere of a light wave with initial polarization $\vec{S_0}$ passing through an absorbing polarizer $\vec{p_1}$, then through a total reflection based polarizer $\vec{p_2}$ and finally through an absorbing polarizer $\vec{p_3}=\vec{S_0}$. We use a stereographic projection of the Poincaré sphere viewed from $-\vec p_2$. The polarization $\vec{p_2}$ is here the center of the stereographic projection. The solid line represents the actual trajectory followed by the polarization state: this trajectory follows the geodesic relying $\vec{S_0}$ to $\vec{p_1}$, a loxodrome from $\vec{p_1}$ to $\vec{p_2}$, and again the geodesic relying $\vec{p_2}$ to $\vec{S_0}$. The dashed line delimits the geodesic triangle whose the vertices are the polarization states $\vec{S_0}$, $\vec{p_1}$ and $\vec{p_2}$. The presented loxodrome is such as $\Delta/\Gamma=10$.
}
\end{center}
\end{figure}

Let us consider a light wave with an initial polarization $\vv{S_0}$ going through successively three polarizers with respective polarization axis $\vv{p_1}$, $\vv{p_2}$, $\vv{p_3}=\vv{S_0}$. 
The trajectory $\mathrm{C}$ of the light wave polarization state on the Poincaré sphere is closed since the polarization returns to its initial value $\vv{S_0}$.
We note $\mathcal{C}$ the trajectory of the light wave state in the Hilbert space.
Using (\ref{proj}),  let us associate to the three polarization axis the three normalized Hilbert states $\ket{1}$, $\ket{2}$, $\ket{0}$ such as  $\rho_{\vv{p_1}}=\ket{1}\bra{1}$,  $\rho_{\vv{p_2}}=\ket{2}\bra{2}$, $\rho_{\vv{p_3}}=\rho_{\vv{S_0}}=\ket{0}\bra{0}$.
From now on we will consider normalized states since the evolution of the light wave intensity does not influence the geometric phase and the Pancharatnam phase.
Let us denote by $\ket{\psi_0}=\ket{0}$ the initial state of the light wave. As stated in Section~\ref{sec:pol}, the state of the light wave after the first polarizer is $\ket{\psi_1}=\ket{1}\scal{1}{0}e^{i\phi_{\mathrm{t}}^{01}}$ where $\phi_{\mathrm{t}}^{01}=\arg\scal{\psi_0}{\psi_1}$ is the total phase between light wave states $\ket{\psi_0}$ and $\ket{\psi_1}$. Also, the light wave state after the second and the third polarizers are respectively
$\ket{\psi_2}=\ket{2}\scal{2}{\psi_1}e^{i\phi_{\mathrm{t}}^{12}}=\ket{2}\scal{2}{1}\scal{1}{0}e^{i\left(\phi_{\mathrm{t}}^{01}+\phi_{\mathrm{t}}^{12}\right)}$ and
$\ket{\psi_3}=\ket{0}\scal{0}{\psi_2}e^{i\phi_{\mathrm{t}}^{23}}=\ket{0}\scal{0}{2}\scal{2}{1}\scal{1}{0}e^{i\left(\phi_{\mathrm{t}}^{01}+\phi_{\mathrm{t}}^{12}+\phi_{\mathrm{t}}^{23}\right)}$
where $\phi_{\mathrm{t}}^{12}=\arg\scal{\psi_1}{\psi_2}$ and $\phi_{\mathrm{t}}^{23}=\arg\scal{\psi_2}{\psi_3}$ are the total phases between respectively the light wave states $\ket{\psi_1}$ and $\ket{\psi_2}$ and the light wave states $\ket{\psi_2}$ and $\ket{\psi_3}$.
As in (\ref{eq:phip}) the total phases $\phi_{\mathrm{t}}^{01}$, $\phi_{\mathrm{t}}^{12}$, and $\phi_{\mathrm{t}}^{23}$ are irrelevant phases which are not related to the three successive polarization processes experienced by the light wave.
Choosing suitable optical paths between polarizers, it is always possible to wipe out the phases $\phi_{\mathrm{t}}^{01}$, $\phi_{\mathrm{t}}^{12}$, and $\phi_{\mathrm{t}}^{23}$
in order to fulfill successively the Pancharatnam \textit{in phase} criterion between light wave states $\ket{\psi_0}$ and $\ket{\psi_1}$, $\ket{\psi_1}$ and $\ket{\psi_2}$, and, 
$\ket{\psi_2}$ and $\ket{\psi_3}$ (see Section~\ref{sec:pancharatnam}). Hence, although two successive light wave states can be said \textit{in phase}, the initial and the final light wave state can not be since
$\phi_{\mathrm{t}}^{03}=
\arg\scal{\psi_0}{\psi_3}=\arg\left(\scal{0}{2}\scal{2}{1}\scal{1}{0}\right)+\phi_{\mathrm{t}}^{01}+\phi_{\mathrm{t}}^{12}+\phi_{\mathrm{t}}^{23}
$. Unlike the phase term $\phi_{\mathrm{t}}^{01}+\phi_{\mathrm{t}}^{12}+\phi_{\mathrm{t}}^{23}$ which depends on the light wave optical path and can be arbitrarily set to zero, the phase term
\begin{equation}
\begin{array}{lll}
\phi_{\mathrm{p}}^{012}&=&\arg\left(\scal{0}{2}\scal{2}{1}\scal{1}{0}\right)\\
&=&\arg\Tr\left(\rho_{\vv{p_2}}\rho_{\vv{p_1}}\rho_{\vv{S_0}}\right),
\end{array}
\end{equation}
named after S. Pancharatnam \cite{pancharatnam56}, is a phase depending only on the relative \textit{loci} of the polarization states on the Poincaré sphere. In the chosen example, the Pancharatnam phase $\phi_{\mathrm{p}}^{012}$ is equal to minus half of the area enclosed by the geodesic triangle whose the vertices correspond to the polarization vectors $\vv{S_0}$, $\vv{p_1}$ and $\vv{p_2}$ on the Poincaré sphere (see the dashed line delimited geodesic triangle on Figure \ref{fig4}).
The interference
of the initial and final light wave states, \textit{i.e.} $\ket{\psi_0}$ and $\ket{\psi_3}$, will give interference fringes which depend on 
the phase $\phi_{\mathrm{t}}^{03}$ and consequently on the Pancharatnam phase $\phi_{\mathrm{p}}^{012}$. As a consequence, changing \textit{e.g.} the first polarizer direction $\vv{p_1}$ to $\vv{p_1}'$ induces a change in the Pancharatnam phase ($\phi_{\mathrm{p}}^{012}\rightarrow\phi_{\mathrm{p}}^{01'2}$) which is detected as a shift of the interference fringes.
In the same time, the phase term $\phi_{\mathrm{t}}^{01}+\phi_{\mathrm{t}}^{12}+\phi_{\mathrm{t}}^{23}$ does not change since $\phi_{\mathrm{t}}^{01}=\phi_{\mathrm{t}}^{01'}$ and $\phi_{\mathrm{t}}^{12}=\phi_{\mathrm{t}}^{1'2}$.
The Pancharatnam phase $\phi_{\mathrm{p}}^{012}$ is a kind of geometric phase which has been measured by interferometry experiments \cite{berry96,hariharan97a,hariharan97b}.

The above presented example is valid for any type of polarizer since the only required assumption is that each intermediate polarization state $\ket{\psi_i}$ is projected on the next one $\ket{\psi_{i+1}}$. 

If all the polarizers crossed by the light wave are absorbing polarizers, the Pancharatnam phase  $\phi_{\mathrm{p}}^{012}$ is indeed the geometric phase acquired by the light wave through the trajectory $\mathcal{C}$, $\phi_{\mathrm{g}}[\mathrm{C}]=\phi_{\mathrm{p}}^{012}$.
Between each intermediate polarization state no geometric phase (\ref{eq:phiap}) is acquired since the polarization state follows a geodesic $\mathrm{C}_{ii+1}$ on the Poincaré sphere, but
along the overall trajectory $\mathrm{C}$, which is obviously not a geodesic since the trajectory is closed, the geometric phase $\phi_{\mathrm{p}}^{012}$ is acquired.


Let us now consider that all the polarizers but the second are absorbing polarizers. We take the second polarizer as a total reflection based polarizer. As stated before the Pancharatnam phase measured by an interferometry experiment will be the same as the one measured in the equivalent experiment using uniquely absorbing polarizers. But the Pancharatnam phase is not the whole geometric phase really acquired by the light wave. Indeed, see Appendix~\ref{app:a}, the geometric phase, in the chosen example, is 
\begin{equation}\label{eq:ppg}
\phi_{\mathrm{g}}[\mathrm{C}]=
\phi_{\mathrm{p}}^{012}
+
\phi_{\mathrm{g}}[\mathrm{C}_{12}].
\end{equation}
Here, the geometric phase $\phi_{\mathrm{g}}[\mathrm{C}]$, which is equal to minus half of the snail shell area presented in Figure 5, is the sum of, 
$\phi_{\mathrm{p}}^{012}$, the Pancharatnam phase, which is related to the area of the $\rho_{\vv{S_0}}$,$\rho_{\vv{p_1}}$,$\rho_{\vv{p_2}}$-geodesic triangle, 
and, $\phi_{\mathrm{g}}[\mathrm{C}_{12}]$, the non zero geometric phase acquired by the light wave during the polarization process between polarization states $\rho_{\vv{p_1}}$, and $\rho_{\vv{p_2}}$, which is related to the complementary area.

More generally, and in even in the quantum realm, let us consider a curve $\mathcal{C}$ in the Hilbert space $\mathcal{H}$ which corresponds to a closed curve $\mathrm{C}$ in the space of rays $\mathcal{R}$. Let us consider that the curve $\mathcal{C}$ is marked out with a set of states $\left\{\ket{\psi_i}\right\}_{i=0,\dots,N}$ such as $\ket{\psi_i}$ is a projection of the previous state $\ket{\psi_{i-1}}$ and such as $\rho_{\vv{S_0}}=\rho_{\vv{S_N}}$. The Pancharatnam phase does not take into account how each successive states $\ket{\psi_i}$ is projected onto the following one, so the Pancharatnam phase $\phi_{\mathrm{p}}^{0\dots N-1}$ will be proportional to the area of the geodesic polygon of which the vertices are the \textit{loci} of the $\rho_{\vv{S_i}}$ states on the Poincaré sphere. \textit{A contrario}, the whole geometric phase $\phi_{\mathrm{g}}[\mathrm{C}]$ includes also the non zero geometric phases $\phi_{\mathrm{g}}[\mathrm{C}_{ii+1}]$ possibly acquired during the successive state 
projections. The whole geometric phase is then
\begin{equation}
\phi_{\mathrm{g}}[\mathrm{C}]=\phi_{\mathrm{p}}^{0\dots N-1}+
\sum_{i=0}^{N-1}\phi_{\mathrm{g}}[\mathrm{C}_{ii+1}].
\end{equation}

As interferometric superposition is sensitive to the total phase $\phi_{\mathrm{t}}^{0N}=\arg\sca{\psi_0}{\psi_N}$, it cannot be used to measure the whole geometric phase acquired by a quantum state. Indeed, if we consider an Hilbert space curve $\mathcal{C}$ of which a point $\ket{\psi_b}\in\mathcal{C}$ is the projection of a previous point $\ket{\psi_a}\in\mathcal{C}$, an interferometry based experiment will be able to measure only the truncate geometric phase $\phi_{\mathrm{g}}[\mathrm{C}]-\phi_{\mathrm{g}}[\mathrm{C}_{ab}]$.
The acquired geometric phase $\phi_{\mathrm{g}}[\mathrm{C}_{ab}]$ between the states $\ket{\psi_a}$ and $\ket{\psi_b}$ is compensated by the non trivial part of the dynamic phase $\phi_{\mathrm{d}}[\mathcal{C}_{ab}]$. This is clearly seen if we compare Eqs.~(\ref{phigf}) and (\ref{eq:phid}). The projection process dependent part of (\ref{eq:phid}) compensates exactly the geometric phase (\ref{phigf}).
The total phase $\phi_{\mathrm{t}}^{0N}$ has no memory of geometric phases possibly acquired between two Hilbert states such as one being the projection of the other.

%
%
%
%

\section{Summary and Conclusion}

In this paper we have considered the geometric phase acquired by a light wave going through elementary polarizing devices. We used the quantum kinematic approach of geometric phases and we have discussed the different cases using area decomposition of the Poincaré sphere. After a review of unitary case, \textit{e.g.} the geometric phase acquired by the light wave going through a birefringent plate or a media with optical activity, we have presented the non unitary case corresponding to the geometric phase acquired by a light wave passing through any polarizer. As expected we retrieve the fact that for ideal absorption based polarizers the light wave gains no geometric phase.
However, for other types of polarizers, such as ideal total reflection based polarizers or realistic polarizers which can be for example dichroic polarizers containing inherently small amount of birefringence, a non trivial geometric
phase is acquired.
This geometric phase is non zero despite the fact that, as seen in (\ref{gpp}),
the initial and the final polarization states are \textit{in phase} according to the Pancharatnam criterion
\cite{pancharatnam56}.
This non zero geometric phase (\ref{eq:phigf}) is indeed acquired by the light wave since the evolution of its normalized Stokes vector does not describe a geodesic of the Poincaré sphere. In the case of total reflection based polarizer, the polarization describes \cite{lages08} on the Poincaré sphere a loxodrome (Figure \ref{fig2}) with a characteristic angle $\chi=\arctan\left(\Gamma/\Delta\right)$ depending on the ratio between the attenuation/absorption rate and the dephasing rate.

In this paper, we have also shown  that the geometric phase acquired by a light wave the initial state of which is projected onto its final state can not be measured using the usual interferometric superposition. As a consequence, interferometry experiments are able to measure the Pancharatnam phase but not the actual complete geometric phase.

\appendix
\section{}
\label{app:a}

We provide here a demonstration of (\ref{eq:ppg}).
Let us compute the geometric phase $\phi_{\mathrm{g}}[\mathrm{C}]$ using the horizontal lift $\mathcal{C}_\mathrm{h}\subset\mathcal{H}$ of the unit
ray space curve $\mathrm{C}\subset\mathcal{R}$. For such a curve $\mathcal{C}_\mathrm{h}$, the dynamic phase vanishes, $\phi_{\mathrm{d}}[\mathcal{C}_\mathrm{h}]=0$, since for any point $\ket{\psi(s)}$ on the curve $\mathcal{C}_\mathrm{h}$, we have $\mathrm{Im}\sca{\psi(s)}{\dot\psi(s)}/\sca{\psi(s)}{\psi(s)}=0$.
As the unit ray space curve $\mathrm{C}$ passes through the polarization states $\rho_{\vv{S_0}}$, $\rho_{\vv{p_1}}$, $\rho_{\vv{p_2}}$, and again $\rho_{\vv{S_0}}$, let us consider that the Hilbert space curve $\mathcal{C}_\mathrm{h}$ passes through the normalized states $\ket{\psi_0}$, $\ket{\psi_1'}\sim\ket{\psi_1}$, 
$\ket{\psi_2'}\sim\ket{\psi_2}$, and
$\ket{\psi_3'}\sim\ket{\psi_0}$.
In such conditions the geometric phase
(\ref{pg}) reads then 
\begin{equation}\label{pgh2}
\phi_{\mathrm{g}}[\mathrm{C}]=\,\arg\sca{\psi_0}{\psi_3'}.
\end{equation}
Let us compute succesively $\ket{\psi_1'}$, $\ket{\psi_2'}$, and $\ket{\psi_3'}$.

\subsection{$\ket{\psi_0}\rightarrow\ket{\psi_1'}$}
The light wave passes through an absorbing polarizer bringing progressively the light wave polarization onto $\vv{p_1}$. The light wave state experiences a transformation
\begin{equation}
\begin{array}{l}
\ket{\psi(s)}=\ds\frac{e^{i\alpha(s)}}{\cosh\gamma_1(s)+\vv{S_0}\cdot\vv{p_1}\sinh\gamma_1(s)}\\
\times e^{\frac{\gamma_1(s)}{2}\vv{p_1}\cdot\vv{\sigma}}\,\ket{\psi_0}
\end{array}
\end{equation}
where the $s$ parameter runs from $s_0$ to $s_1$ with $\ket{\psi(s_0)}=\ket{\psi_0}$ and
$\ket{\psi(s_1)}=\ket{\psi_1'}$.
The absorption parameter $\gamma_1(s)$ is such as $\gamma_1(s_0)=0$ and $\gamma_1(s_1)\rightarrow+\infty$.
The gauge parameter $\alpha(s)$ is such as $\alpha(s_0)=0$ and such as $\mathrm{Im}\sca{\psi(s)}{\dot\psi(s)}=0$ for all $s\in[s_0,s_1]$. A straightforward calculus gives $\alpha(s)=\alpha(s_0)=0$ and
\begin{equation}
\ket{\psi_1'}=\ds\frac{2}{1+\vv{S_0}\cdot\vv{p_1}}\,\rho_{\vv{p_1}}\,\ket{\psi_0}.
\end{equation}
As expected, no geometric phase is acquired by the light wave during this transformation since $\arg\sca{\psi_0}{\psi_1'}=0$.

\subsection{$\ket{\psi_1'}\rightarrow\ket{\psi_2'}$}
The light wave passes now through a total reflection based polarizer bringing progressively the light wave polarization onto $\vv{p_2}$. The light wave state experiences a transformation
\begin{equation}
\begin{array}{l}
\ket{\psi(s)}=\ds\frac{e^{i\alpha(s)}}{\cosh\gamma_2(s)+\vv{p_1}\cdot\vv{p_2}\sinh\gamma_2(s)}\\
\times e^{\frac{\gamma_2(s)-i\delta_2(s)}{2}\vv{p_2}\cdot\vv{\sigma}}\,\ket{\psi_1'}
\end{array}
\end{equation}
where the $s$ parameter runs from $s_1$ to $s_2$ with $\ket{\psi(s_1)}=\ket{\psi_1'}$ and
$\ket{\psi(s_2)}=\ket{\psi_2'}$.
The attenuation parameter $\gamma_2(s)$ and the dephasing parameter $\delta_2(s)$ are such as the ratio $\dot\gamma_2(s)/\dot\delta_2(s)$ is finite (see Section~\ref{trbp}) with $\gamma_2(s_1)=\delta_2(s_1)=0$ and $\gamma_2(s_2)\rightarrow+\infty$.
The gauge parameter $\alpha(s)$ is such as $\alpha(s_1)=0$ and such as $\mathrm{Im}\sca{\psi(s)}{\dot\psi(s)}=0$ for all $s\in[s_1,s_2]$. Using (\ref{eq2}), the final state of this intermediate transformation is
\begin{equation}
\ket{\psi_2'}=\ds\frac{2e^{-i\frac{\Omega}{2}}}{1+\vv{p_1}\cdot\vv{p_2}}\,\rho_{\vv{p_2}}\,\ket{\psi_1'}
\end{equation}
where the solid angle $\Omega$ is the area
swept on the Poincaré sphere by the arc length joining the endpoint of the Stokes vector $\vv S(z)$ during the transformation
and the endpoint of the polarization vector $\vv{p_2}$ (see Figure \ref{fig4}). As in (\ref{phigf}), the geometric phase gained by the light wave during this transformation is $\arg\sca{\psi_1'}{\psi_2'}=-\frac{\Omega}{2}$. 

\subsection{$\ket{\psi_2'}\rightarrow\ket{\psi_3'}$}
Finally, the light wave passes through an absorbing polarizer bringing progressively the light wave polarization onto its initial value $\vv{S_0}$. The light wave state experiences a transformation
\begin{equation}
\begin{array}{l}
\ket{\psi(s)}=\ds\frac{e^{i\alpha(s)}}{\cosh\gamma_3(s)+\vv{p_2}\cdot\vv{S_0}\sinh\gamma_3(s)}\\
\times e^{\frac{\gamma_3(s)}{2}\vv{S_0}\cdot\vv{\sigma}}\,\ket{\psi_2'}
\end{array}
\end{equation}
where the $s$ parameter runs from $s_2$ to $s_3$ with $\ket{\psi(s_2)}=\ket{\psi_2'}$ and
$\ket{\psi(s_3)}=\ket{\psi_3'}$.
The absorption parameter $\gamma_3(s)$ is such as $\gamma_3(s_2)=0$ and $\gamma_3(s_3)\rightarrow+\infty$.
The gauge parameter $\alpha(s)$ is such as $\alpha(s_2)=0$ and such as $\mathrm{Im}\sca{\psi(s)}{\dot\psi(s)}=0$ for all $s\in[s_2,s_3]$. As in the $\ket{\psi_0}\rightarrow\ket{\psi_1'}$ transformation, a straightforward calculus gives $\alpha(s)=\alpha(s_2)=0$ and
\begin{equation}
\ket{\psi_3'}=\ds\frac{2}{1+\vv{p_2}\cdot\vv{S_0}}\,\rho_{\vv{S_0}}\,\ket{\psi_2'}.
\end{equation}
Again, as expected, no geometric phase is acquired by the light wave during this transformation since $\arg\sca{\psi_2'}{\psi_3'}=0$.

\subsection{$\ket{\psi_0}\rightarrow\ket{\psi_1'}\rightarrow\ket{\psi_2'}\rightarrow\ket{\psi_3'}$}

The geometric phase acquired through the complete transformation is then
\begin{equation}
\begin{array}{lll}
\phi_{\mathrm{g}}[\mathrm{C}]&=&\,\arg\sca{\psi_0}{\psi_3'}\\
&=&-\ds\frac\Omega2+\arg\bra{\psi_0}\rho_{\vv{S_0}}
\,\rho_{\vv{p_2}}
\,\rho_{\vv{p_1}}\,\ket{\psi_0}\\
&=&-\ds\frac\Omega2+\arg\Tr
\left(
\rho_{\vv{p_2}}
\,\rho_{\vv{p_1}}
\,\rho_{\vv{S_0}}
\right)\\
&=&\phi_{\mathrm{g}}[\mathrm{C_{12}}]+\phi_{\mathrm{p}}^{012}.
\end{array}
\end{equation}












\end{document}